\documentclass[a4paper,11pt]{article}
\pdfoutput=1 
\usepackage{jheppub} 
                     
\usepackage{booktabs,makecell, bm}
\usepackage[font=small]{caption}
\usepackage{framed}
\usepackage{hyperref}
\usepackage[export]{adjustbox}
\usepackage{amsmath}
\usepackage{lineno}
\usepackage{ytableau}
\usepackage{subfig}
\usepackage{url}
\usepackage[numbers]{natbib}
\usepackage[colorinlistoftodos]{todonotes}

\usepackage{xcolor}

\renewcommand{\[}{\begin{equation}}
\renewcommand{\]}{\end{equation}}

\makeatother

\newcommand{\comm}[2]{[#1,#2]}                              

\global\long\global\long\def\Tr{\mbox{Tr}}

\newcommand{\acomm}[2]{\{#1,#2\}}

\def\bea{\begin{eqnarray}}
\def\eea{\end{eqnarray}}
\def\be{\begin{equation}}
\def\ee{\end{equation}}
\def\ea{\end{align}}
\def\bse{\begin{subequations}}
\def\ese{\end{subequations}}
\def\1F1{{}_1\!F_1}
\def\2F0{{}_2\!F_0}

\title{BMN-like Matrix Models}

\author{Eunwoo Lee}

\affiliation{Department of Theoretical Physics, \\ Tata Institute
	of Fundamental Research, Homi Bhabha Rd, Mumbai 400005, India}
\emailAdd{eunwoo.lee@tifr.res.in}

\abstract{We conjecture a family of matrix quantum mechanical models that are holographically dual to discrete light-cone quantization of M-theory in pp-wave–like backgrounds. These backgrounds can be obtained from a Penrose limit of AdS$_4\times X_7$, where $X_7$ is Einstein. The matrix models arise from a classically consistent dimensional reduction of the UV Lagrangians of $\mathcal{N}=1$ superconformal field theories, in close analogy with how the BMN matrix model is obtained by dimensional reduction from $\mathcal{N}=4$ super Yang–Mills theory. We also discuss about supersymmetric black objects in pp-wave background by studying the Witten index and speculate that the area of the horizon is bounded from above for a fixed $N$.
}

\begin{document} 

\maketitle

\section{Introduction}

The BMN matrix model \cite{Berenstein:2002jq} -- also known as the mass-deformed BFSS matrix model \cite{Banks:1996vh} -- is a mass deformation of the BFSS matrix quantum mechanics describing the discrete light-cone quantization (DLCQ) of M-theory in the maximally supersymmetric eleven-dimensional pp-wave background \cite{Kowalski-Glikman:1984qtj,Dasgupta:2002hx,Dasgupta:2002ru,Michelson:2002wa,Kim:2002if} (at large $N$, large $R$, large coupling). The metric of the background is
\begin{align}\label{metric 1}
    ds^2 &= -2\,dx^{+}dx^{-} + \sum_{I=1}^{9}(dx_I)^2- \left[ \sum_{a=1}^{3}\frac{\mu^2}{9}x_a^2 +\sum_{i=4}^{9}\frac{\mu^2}{36}x_i^2 \right](dx^{+})^2
\end{align} 
with the RR field strength $F_{+123}=\mu$. It can be obtained as the Penrose limit of AdS$_4\times S^7$ or AdS$_7\times S^4$ \cite{Blau:2002dy,Dasgupta:2002hx}. The BMN matrix model also arises as the worldvolume theory of $N$ D0-branes in the reduced type IIA background, with mass terms and cubic couplings induced by the background flux \cite{Dasgupta:2002hx}. This provides a quantum-mechanical model with a rich nonperturbative structure, whose spectrum and supersymmetry properties have been studied \cite{Kim:2002if,Dasgupta:2002ru}. 
\footnote{%
The term “BMN model” is sometimes also used to denote the BMN limit of $\mathcal{N}=4$ super Yang–Mills theory, which isolates a sector of nearly BPS operators dual to string excitations in the ten-dimensional pp-wave background obtained as the Penrose limit of AdS$_5\times S^5$. In this paper, however, the term “BMN matrix model” refers exclusively to the mass-deformed BFSS matrix quantum mechanics describing the DLCQ of M-theory in the eleven-dimensional pp-wave background.
} 

There are several reasons why the BMN quantum mechanics continues to be of interest. First, it provides a rare example of a fully supersymmetric, massive, and yet nontrivial quantum mechanical system with an M-theory interpretation. The mass deformation lifts all classical flat directions of the BFSS potential, stabilizing the vacuum and allowing for controlled studies of strongly coupled dynamics. At the same time, supersymmetry remains intact, making it possible to obtain certain exact results. These include the identification of supersymmetric vacua with bubbling geometries \cite{Lin:2004kw,Lin:2005nh,Kawahara:2006hs}, as well as exact results for correlation functions derived using supersymmetric localization \cite{Asano:2012zt,Asano:2014vba}.

In addition, BMN and, more generally, matrix models are considerably more amenable to numerical simulation than higher-dimensional quantum field theories \cite{Catterall:2007fp,Anagnostopoulos:2007fw,Catterall:2008yz,Catterall:2010gf,Hanada:2011fq,Costa:2014wya,Gur-Ari:2015rcq,Hanada:2016zxj,Maldacena:2023acv,Komatsu:2024vnb}. Recent developments in numerical methods, bootstrap techniques, and holographic approaches have opened new nonperturbative avenues for studying these systems. Progress in high-precision simulations and bootstrap-style techniques for low-dimensional and matrix quantum mechanics has led to renewed interest in studying D0-brane and related matrix models \cite{Lin:2025srf}.
For broader perspectives and recent developments, we refer the reader to the recent reviews \cite{Lin:2025iir,Fliss:2025omb}, which provide comprehensive overviews of matrix models and their applications.

\medskip

An interesting structural fact of the BMN matrix model is that it can be obtained from a spherical Kaluza–Klein reduction of $U(N)$ $\mathcal{N}=4$ SYM: keeping only the lowest spherical harmonic modes on $S^3$ that preserve an $SU(2)_L\subset SO(4)$ subgroup yields the mass terms and interaction structure of the BMN deformation at the classical level \cite{Kim:2003rza}.
In other words, the BMN model appears as a consistent truncation (at the classical level) of the four-dimensional parent theory when only the lowest harmonics are retained. 
Regarding $\mathcal{N}=4$ SYM being directly related to $10d$ IIB theory, it is surprising—and not very well understood—that its truncation is related to the physics of $11d$ M-theory (or IIA in the weak coupling regime).

These facts lead to two natural questions that this paper addresses:
First, what happens if one carries out an analogous KK reduction on $S^3$ for other four-dimensional supersymmetric quantum field theories, especially those with known holographic duals with UV Lagrangian descriptions (for example orbifolds of $\mathcal{N}=4$, the Klebanov–Witten theory on $T^{1,1}$, the $Y^{p,q}$ families, or exactly marginal $\beta$-deformations of $\mathcal{N}=4$ SYM \cite{Klebanov:1998hh,Martelli:2004wu,Benvenuti:2004dy,Kachru:1998ys,Lawrence:1998ja})? Is there a truncation that preserves the classical equations of motion and yields a one-dimensional supersymmetric matrix quantum mechanics? Following the methods of \cite{Kim:2003rza,Asplund:2015yda}, we show that such reductions are possible for broad classes of holographic 4d theories and identify the resulting quantum-mechanical models.

Second, do the resulting matrix quantum mechanics possess holographic dual descriptions analogous to the BMN correspondence? Put differently, can one identify dual M-theory backgrounds that capture the dynamics of these reduced matrix models, and what features—such as fluxes, isometries, and preserved supersymmetries—distinguish different background geometries? 
Here, we suggest that there exist such dual geometries. 

Let us briefly describe a simple geometric ansatz that descibres the background geometry. First, consider a 4d theory which is dual to IIB string theory in AdS$_5\times Y_5$ and perform a truncation to obtain a matrix quantum mechanics.
We claim that the matrix model is dual to the DLCQ of M-theory in the \emph{pp-wave--like geometry}, where we replace the ``internal'' $S^5$ in the pp-wave geometry with a general 5-dimensional internal manifold $Y_5$ appearing in the holographic theory. To illustrate this, we rewrite the metric of the pp-wave geometry \eqref{metric 1} dual to the BMN model as
\begin{align} ds^2 &= -2\,dx^{+}dx^{-} + \sum_{I=1}^{3}(dx_a)^2+dr^2+r^2 d\Omega_5^2 - \left[ \sum_{a=1}^{3}\frac{\mu^2}{9}x_a^2 +\frac{\mu^2}{36}r^2 \right](dx^{+})^2, \end{align}
with the RR field strength $F_{+123} = \mu$. Now consider a class of $U(N)$ matrix models obtained via Kaluza–Klein reduction from holographic theories dual to type IIB string theory on AdS$_5 \times Y_5$. We conjecture that the corresponding pp-wave-like background takes the form
\begin{align}\label{mod geo}
ds^2 &= -2\,dx^{+}dx^{-} + \sum_{I=1}^{3}(dx_a)^2+dr^2+r^2 dY_5^2 - \left[ \sum_{a=1}^{3}\frac{\mu^2}{9}x_a^2 +\frac{\mu^2}{36}r^2 \right](dx^{+})^2, \end{align}
where $dY_5^2$ denotes the metric on $Y_5$.

In Subsection \ref{holo}, we present examples of this construction and check consistency by examining the symmetry structure of the resulting matrix quantum mechanics and the associated pp-wave–like geometry. We show that \eqref{mod geo} satisfies the Einstein equations, and verify that the number of Killing spinors matches those in the matrix quantum mechanics.

Turning off the mass term proportional to \(\mu\), the matrix model reduces to a BFSS-like theory. At weak coupling this model describes D0-branes in type IIA string theory propagating on a background of the form \(\mathbb{R}^{1,3}\times X_6\), where \(X_6\) is a complex three-dimensional space (e.g. non-compact Calabi–Yau) whose base is the five-dimensional manifold \(Y_5\) (e.g. Sasaki-Einstein manifold). From this perspective, the BMN-like matrix models can be viewed as mass deformations of the corresponding BFSS-like models.
Such matrix models, describing D-branes probing Calabi–Yau singularities, have been extensively studied since the pioneering work of \cite{Douglas:1996sw}. A more detailed discussion is given in Subsection~\ref{subsec:D0_on_CY3}.

\medskip
As a brief but important aside, in Subsection \ref{susy bh} we study a supersymmetric deconfined phase of the BMN matrix model. This phase can be interpreted as being dual to a supersymmetric black object in a plane-wave background, with an entropy scaling as $O(N^2)$ when the charges also scales as $O(N^2)$.
Although we were not able to compute the entropy of the matrix model exactly, we place an upper bound by noting that the entropy of this supersymmetric black object is bounded above by the free BPS partition function. Exploiting the fact that the BMN matrix model is a multi-matrix model, one can show that the \textit{free} BPS partition function grows at most logarithmically with the charges, with a logarithmic coefficient proportional to $N^{2}$.

Consequently, the entropy of the index $S_{\rm ind}$(and similarly, of the BPS partition function at finite coupling $S_{\rm BPS}$) is bounded from above by a logarithmic function of the charge, and can be written as
\begin{align}
S_{\rm ind}<S_{\rm BPS} < S_{\rm free ~BPS}\sim b_1 N^2 + b_0 N^2 \log(Q/N^2), \qquad Q \gg O(N^2),
\end{align}
where $b_0$ and $b_1$ are constants independent of the charges.

We further speculate that the actual entropy growth of the index in the regime $Q \gg O(N^2)$ should instead take the form
\begin{align}
S_{\rm ind} \sim c_1 N^2 + c_0 N^{\alpha} \log(Q/N^2),
\end{align}
with $\alpha<2$ and $c_0, c_1$ constants \emph{independent} of the charges. This behavior suggests that the horizon area of the black object, measured in units of $N^2$, is saturated by the constant term $c_1$, while the logarithmic contribution may be interpreted as arising from stringy excitations around a black object of maximal size.

This stands in clear contrast to supersymmetric black holes in AdS, whose horizon area can grow without bound as the charges are increased. We return to a more detailed discussion of these points in Subsection \ref{susy bh}.

\medskip
 
The paper is organized as follows. In Section \ref{review} we review the BMN matrix model and its M-theory dual. We describe how the KK reduction procedure of $\mathcal{N}=4$ SYM on $S^3\times\mathbb{R}^1$ yields the BMN matrix model.
In Subsection \ref{susy bh}, we take a short digression to discuss about supersymmetric black holes in pp-wave background.
In Section~\ref{bmnlike}, we derive matrix quantum mechanics via dimensional reduction of several families of four-dimensional parent theories (orbifolds of $\mathcal{N}=4$, the Klebanov–Witten theory, $Y^{p,q}$ families, and marginal $\beta$-deformations) and analyze their symmetry and BPS structure. We propose dual background geometries. 
We also comment on truncations of non-holographic supersymmetric field theories and truncations of $\mathcal{N}=4$ $SO,Sp$ gauge groups.
In Section \ref{discussion} we end with a discussion and directions for future work.

\section{BMN matrix model}\label{review}

In this section, we review the BMN matrix model \cite{Berenstein:2002jq,Dasgupta:2002hx} and its realization as a Kaluza–Klein reduction of $\mathcal{N}=4$ super Yang–Mills theory. Subsections \ref{pp wave back}, \ref{bmn from sym}, and \ref{sym class} are largely review material, following in particular \cite{Kim:2003rza,Dasgupta:2002hx} and references therein.
Although this section is mostly a review, it also contains some new results and viewpoints.

The BMN matrix model can be regarded as a massive deformation of the BFSS matrix model that preserves maximal supersymmetry. The BMN matrix model provides a nonperturbative description of M-theory in the light-cone frame on a maximally supersymmetric pp-wave background \cite{Berenstein:2002jq}. Its action can be written as (following the convention of \cite{Kim:2003rza})
\begin{align}\label{bmn lag}
S_\text{flat} & =  \int\!dt\: \Tr \Bigl[
  \tfrac{1}{2} (D_t X_I)^2
  - i \theta D_t \theta
  + \tfrac{1}{4} \comm{X_I}{X_J}^2
  + \theta \Gamma^I \comm{X_I}{\theta} \Bigr] , \nonumber\\
S_M & =  \int\!dt\: \Tr \Bigl[
  - \tfrac{1}{2} \bigl(\tfrac{m}{3}\bigr)^2 (X_a)^2
  - \tfrac{1}{2} \bigl(\tfrac{m}{6}\bigr)^2 (X_i)^2
  + \tfrac{m}{4} i\, \theta \Gamma_{123} \theta
  + \tfrac{m}{3} i\, \epsilon_{abc}\, X_a X_b X_c \Bigr] .
\end{align}

Here, $X$ denotes a scalar in adjoint representation of $U(N)$. The transverse $SO(9)$ index $I=1,\ldots,9$ is split into two subsets, $a=1,2,3$ and $i=4,\ldots,9$, corresponding to the $SO(3)\times SO(6)$ symmetry decomposition. $\theta^A_{\alpha}$ denotes an adjoint-valued fermion, where $A=1,\cdots,4$.
The parameter $m$ is dimensionless and defined as
$$
m = \frac{\mu \alpha'}{2R},
$$
where $R$ denotes the radius of the compact eleventh dimension. The covariant derivative acts as
$$
D_t {\cal O} = \partial_t {\cal O} - i \comm{\omega}{{\cal O}},
$$
with $\omega$ being the gauge field. In the limit $m \to 0$, the mass and cubic terms vanish, and one recovers the BFSS matrix model describing M-theory in flat space.

\subsection{The pp-wave background}\label{pp wave back}

The BMN matrix model is conjectured to be dual to M-theory on a maximally supersymmetric pp-wave background. The metric takes the form
\begin{align}\label{ppmetric}
    ds^2 = -2\,dx^{+}dx^{-} 
    + \sum_{I=1}^{9}(dx^I)^2
    - \left[
      \sum_{a=1}^{3}\frac{\mu^2}{9}x_a^2
      + \sum_{i=4}^{9}\frac{\mu^2}{36}x_i^2
      \right](dx^{+})^2 ,
\end{align}
and the only non-vanishing component of the four-form flux is $F_{+123} = \mu.$ Also, the light-cone momentum is quantized as $P_{-}=\frac{N}{R}$ where $x^-\sim x^-+2\pi R$.
This background preserves 32 supercharges and thus represents one of the maximally supersymmetric solutions of eleven-dimensional supergravity \cite{Kowalski-Glikman:1984qtj}. The geometry can be obtained as a Penrose limit of either $\text{AdS}_4 \times S^7$ or $\text{AdS}_7 \times S^4$ \cite{Blau:2002dy,Berenstein:2002jq}. 

\subsubsection{Killing spinor equation}
We review the fact that the Killing spinor equations admit 32 independent solutions. The Killing spinor equation is written as follows.
\begin{align}
\nabla_M \epsilon + \frac{1}{288}\left(
\Gamma_M^{\;\;NPQR} - 8 \delta_M^N \Gamma^{PQR}
\right) F_{NPQR} \epsilon = 0.
\end{align}
We take the standard Brinkmann-frame vielbein
\begin{align}
e^+ = dx^+, \qquad
e^- = dx^- + \tfrac12 H(x^I)dx^+, \qquad
e^I = dx^I,
\end{align}
where
\begin{align}
H(x) = \sum_a\frac{\mu^2}{9} x_a^2 + \sum_i\frac{\mu^2}{36} x_i^2.
\end{align}
The nonzero spin connection one-forms are given by
\[
\omega^{-I} = \tfrac12 \partial_I H\, e^+,
\qquad
\text{that is,} \quad
\omega^{-a} = \frac{\mu^2}{9} x_a e^+,\quad
\omega^{-i}=\frac{\mu^2}{36} x_i e^+.
\]
All other components of $\omega^{AB}$ vanish.  
Equivalently, in coordinate components,
\[
\omega_{+}^{\ -I} = \tfrac12\partial_I H.
\]

Let us now write out the Killing spinor equations explicitly.  
For the transverse directions ($I=1,\ldots,9$),
\begin{align}\label{trans}
\partial_I \epsilon
= -\,\frac{\mu}{12}
\Gamma_{I}^{+123}
\epsilon.
\end{align}
For $M=\pm$, we find
\begin{align}
\partial_+ \epsilon
= -\,\frac{1}{4}(\partial_I H)\Gamma_{-I}\epsilon
+\frac{\mu}{12}
\Gamma^{123}\epsilon,
\qquad 
\partial_- \epsilon = 0.
\end{align}

The general solution can be classified into two types:
\begin{enumerate}
\item \textbf{Kinematical Killing spinors}, satisfying $\Gamma^+\epsilon=0$.  
  For these spinors, $\partial_I\epsilon=\partial_-\epsilon=0$, leading to
  \begin{align}
  \epsilon = e^{\frac{\mu}{12}\,x^+\,\Gamma^{123}}\,\epsilon_0,
  \quad \Gamma^+\epsilon_0=0,
  \end{align}
  corresponding to $16$ independent solutions.
\item \textbf{Dynamical Killing spinors}, for which $\Gamma^+\epsilon \neq 0$.  
  These take the form
  \begin{align}
  \epsilon =
  \Bigl(
  1 - \frac{\mu}{12}\,x^I \Gamma_{I}^{+123}\Bigr)\chi(x^+),
  \end{align}
  where $\chi$ depends only on $x^+$. This ansatz satisfies the equation \eqref{trans} in directions transverse to $x^+$. The remaining $+$ component then fixes $\chi$, yielding an additional 16 solutions \cite{Figueroa-OFarrill:2001hal,Gauntlett:2002cs}.
\end{enumerate}
Together, these constitute the full set of 32 Killing spinors, confirming that the pp-wave background is maximally supersymmetric.

\subsubsection{Killing spinor equation in ``spherical" coordinates}
For future reference, it is useful to solve the Killing spinor equation after rewriting the pp-wave metric in the following ``spherical" coordinates:
\begin{align}\label{sphere ppwave}
    ds^2=-2dx^{+}dx^{-} +\sum_{a=1}^{3}(dx^a)^2+dr^2+r^2d\Omega_5^2-\left[\sum_{a=1}^{3}\frac{\mu^2}{9}x_a^2+\frac{\mu^2}{36}r^2\right](dx^+)^2,
\end{align}
with $F_{+123}=\mu$. 
We have used $(dx^i)^2=dr^2+r^2d\Omega_5^2$ and $\sum_i(x^{i})^2=r^2$.

One can work in an orthonormal frame,
\begin{align}
    e^r = dr,\qquad e^m = r\,\tilde e^m(\theta),
\end{align}
where $\tilde e^m$ denotes an orthonormal vielbein on the unit five-sphere and indices $m,n,\dots$ label tangent-frame directions on $S^5$.
The nonzero spin connection components are
\begin{align}
    \omega^{-a}=\frac{\mu^2}{9}x_a e^+,\quad
    \omega^{-r}=\frac{\mu^2}{36}r e^+,\quad
    \omega_{m}^{rn}=\tilde{e}^n_m,\quad 
    \tilde{\omega}_{m}^{np},
\end{align}
where $\widetilde\omega$ is the intrinsic connection on the unit sphere.
Therefore, the Killing spinor equations in transverse directions that correspond to \eqref{trans} become
\begin{align}\label{kispin}
\nabla_r \epsilon + \frac{\mu}{12}\, \Gamma_r^{+123} \epsilon = 0,
\qquad
\nabla_m \epsilon + \frac{\mu}{12}\, \Gamma_m^{+123} \epsilon = 0.
\end{align}
To solve these, we adopt the ansatz
\begin{align}
\epsilon(r,\theta) = S(\theta)\, \epsilon_r(r),
\end{align}
with
\begin{align}
\epsilon_r(r) = \left( 1 - \frac{\mu}{12} r\, \Gamma_r^{+123} \right) \epsilon_0. 
\end{align}
$\epsilon$ is independent of $r$. The radial equation is then trivially satisfied:
\begin{align}
\nabla_r \epsilon + \frac{\mu}{12} \Gamma_r^{+123} \epsilon 
= (\partial_r \epsilon_r + \frac{\mu}{12} \Gamma_r^{+123} \epsilon_r)S(\theta) = 0.
\end{align}
For the angular directions $m$, we obtain
\begin{align}\label{killing}
\nabla_m \epsilon 
= (\partial_m S)\, \epsilon_r + S\, \partial_m \epsilon_r 
+ \frac{1}{4} \omega_m^{AB} \Gamma_{AB} S \epsilon_r.
\end{align}
Using $\partial_m\Gamma_r = \frac{1}{r}\Gamma_m$\footnote{For instance, in two dimensions, $\Gamma_r=\cos\theta\Gamma_x+\sin\theta\Gamma_y$, giving $\partial_\theta \Gamma_r = \frac{1}{r}\Gamma_{\theta}$.}, the Killing spinor equation
\begin{align}
\nabla_{m}\epsilon=-\frac{\mu}{12}\Gamma_{m}^{+123}\epsilon=S\left(-\frac{\mu}{12}\Gamma_m^{+123}\right)\epsilon_0,
\end{align}
can be rewritten as
\begin{align}\label{kispin2}
    \nabla_m S = \partial_m S + \frac{1}{4} \tilde{\omega}_m^{pq} \Gamma_{pq} S - \frac{1}{2r} \Gamma_{mr} S
    = \tilde{\nabla}_m S - \frac{1}{2r}\Gamma_{mr}=0,
\end{align}
where $\tilde{\nabla}$ denotes the intrinsic covariant derivative on the unit sphere.
The existence of such an $S(\theta)$ is guaranteed by the Killing spinor structure on the five-sphere, since \eqref{kispin2} is equivalent to solving the Killing spinor equation on the five sphere.

\subsection{BMN matrix model from $\mathcal{N}=4$ SYM}\label{bmn from sym}

Interestingly, the BMN matrix model can be derived from the Kaluza--Klein (KK) reduction of $\mathcal{N}=4$ super Yang--Mills theory compactified on $S^3$ \cite{Kim:2003rza}. The degrees of freedom of $\mathcal{N}=4$ SYM in the lowest KK sector reduce to a set of matrix variables that reproduce the BMN matrix model Lagrangian.

The action of $\mathcal{N}=4$ SYM on a curved background with metric $g_{\mu\nu}$ takes the form
\begin{align}
S = \frac{2}{g_{\text{YM}}^2} \int d^4x \sqrt{|g|}\;
\mathrm{tr}\Big[
  &-\frac{1}{4} F^{\mu\nu} F_{\mu\nu}
  -\frac{1}{2} D^{\mu} \phi_i D_{\mu} \phi_i
  -\frac{\mathcal{R}}{12}\, \phi_i^2
  +\frac{1}{4} [\phi_i, \phi_j]^2
  -2i\, \lambda_A^{\dagger} \sigma^{\mu} D_{\mu} \lambda^A
  \nonumber\\
  &+ (\rho_i)^{AB}\, \lambda_A^{\dagger} i\sigma^2 [\phi_i, \lambda_B^*]
  - (\rho_i)_{AB}\, (\lambda_A^{\dagger})^{\top} i\sigma^2 [\phi_i, \lambda^B]
\Big] .
\end{align}
Here, 
the gauge-covariant derivative acts as $D_{\mu} = \nabla_{\mu} - i [A_{\mu},\, \cdot\,],$
where $\nabla_{\mu}$ denotes the space-time covariant derivative. The scalar fields $\phi_i$ ($i=1,\ldots,6$) transform in the adjoint representation, and the fermions $\lambda_A$ ($A=1,\ldots,4$) carry $SU(4)$ indices. The matrices $(\rho_i)^{AB}$ are Clebsch–Gordan coefficients that couple two $\mathbf{4}$’s of $SU(4)$ into the antisymmetric representation $\mathbf{6}$ corresponding to the six real scalars.

The equation of motion is written as 
\begin{align}\label{full eq}
& D^\nu F_{\nu\mu} + i \comm{\phi_i}{D_\mu \phi_i}
  + 2 \acomm{ \lambda^\dagger }{ \sigma_\mu \lambda }
  = 0  \nonumber\\
& D^\mu D_\mu \phi_i
  - \tfrac{1}{R^2} \phi_i
  + \comm{ \phi_j }{ \comm{ \phi_i }{ \phi_j } }
  - \acomm{ \lambda^\dagger }{ i\sigma^2 \rho_i \lambda^* }
  + \acomm{ \lambda^\top }{ i \sigma^2 \rho_i^\dagger \lambda }
  = 0  \\
& i \sigma^\mu D_\mu \lambda
  - i \sigma^2 \rho_i \comm{\phi_i}{\lambda^*}
  = 0 \nonumber.
\end{align}
Prior to reduction, the fields are expanded in scalar, spinor, and vector spherical harmonics on $S^3$. Working in the Coulomb gauge $\nabla_a A^a = 0,$ where $a$ denotes the spatial directions,
the field expansions take the form \cite{Kim:2003rza}
\begin{align}
\phi_i(x)            & = \sum_{k=0}^\infty \sum_{n=1}^{(k+1)^2}    \phi_i^{kn}(t) Y_{(0)}^{kn}(x) , \\
\lambda_\alpha^A(x)  & = \sum_{k=0}^\infty \sum_{n=1}^{(k+1)(k+2)} \sum_\pm \lambda^{A,kn\pm}(t) Y_{(1/2)\:\alpha}^{kn\pm}(x) , \\
A_0(x)               & = \sum_{k=0}^\infty \sum_{n=1}^{(k+1)^2}    \omega^{kn}(t) Y_{(0)}^{kn}(x) , \\
A_a(x)               & = \sum_{k=0}^\infty \sum_{n=1}^{(k+1)(k+3)} \sum_\pm A^{kn\pm}(t) Y_{(1)\:a}^{kn\pm}(x) .
\end{align}
Here, $Y_{(0)}$, $Y_{(1/2)\,\alpha}$, and $Y_{(1)\,a}$ denote the scalar, spinor, and vector spherical harmonics on $S^3$, respectively. The integer $k=0,1,2,\ldots$ labels different irreducible representations of $SU(2)$, and the index $n$ enumerates the elements within each representation, running from $1$ to its dimension.

In the KK reduction, we keep only the lowest ($k=0$) modes of $A^{kn+}$, $\lambda^{A,kn+}$, and $\phi^{kn}$. The truncation ansatz is therefore
\begin{align}\label{ansatz}
  \phi_i(x) & = X_i (t), \nonumber\\
  \lambda^A_{\alpha}(x) & = \sum_{\hat{\alpha}=1}^2 \theta^A_{\hat{\alpha}} (t) S_\alpha^{\hat{\alpha}+}(x), \nonumber\\
  A_0(x) & = \omega (t), \nonumber\\
  A_a(x) & = \sum_{\hat{a}=1}^3 X_{\hat{a}} (t) V_a^{\hat{a}+}(x),
\end{align}
where $S_\alpha^{\hat{\alpha}+}(x)$ and $V_a^{\hat{a}+}(x)$ are the normalized Killing spinors and Killing vectors on $S^3$. This truncation preserves the lowest harmonic modes and keeps only the singlets of $SU(2)_R\subset SO(4)\simeq SU(2)_L\times SU(2)_R$. This is referred to as the \emph{BMN truncation}.

To verify the consistency of this truncation, we substitute the ansatz \eqref{ansatz} into \eqref{full eq}. Remarkably, all $S^3$-dependent terms match among the equations, ensuring that the truncated system forms a closed set. For instance, the second equation of motion in \eqref{full eq} is independent of the angular coordinates on $S^3$, while the last one is proportional to the spinor harmonics $S_\alpha^{\hat{\alpha}}(x)$. The one-dimensional fields $X_I(t)$, $\omega(t)$, and $\theta(t)$ obey the reduced equation of motion:
\begin{align}\label{bmn eq}
& \comm{X_I}{i D_t X_I}
  - 2 \acomm{ \theta^{\dagger} }{ \theta }
  = 0 , \nonumber \\
& D^2_t X_{\hat{a}}
  + \tfrac{4}{R^2} X_{\hat{a}}
  - \tfrac{6i}{R} \epsilon_{\hat{a}\hat{b}\hat{c}} X_{\hat{b}}X_{\hat{c}}
  - \comm{ X_I }{ \comm{ X_{\hat{a}} }{ X_I } }
  - 2 \acomm{ \theta^{\dagger} }{ \sigma_{\hat{a}} \theta }
  = 0 , \nonumber \\
& D^2_t X_i
  + \tfrac{1}{R^2} X_i
  - \comm{ X_I }{ \comm{ X_i }{ X_I } }
  + \acomm{ \theta^{\dagger} }{ i \sigma^2 \rho_i \theta^{*} }
  - \acomm{ \theta^\top }{ i \sigma_2 \rho_i^\dagger \theta }
  = 0 , \nonumber \\
& i D_t \theta
  - \tfrac{3}{2R} \theta
  + \comm{ X_{\hat{a}} }{ \sigma_{\hat{a}} \theta }
  - \comm{ X_i }{ i \sigma_2 \rho_i \theta^* }
  = 0 .
\end{align}
These are precisely the equations of motion derived from the BMN matrix model Lagrangian \eqref{bmn lag} upon the identification \cite{Kim:2003rza}
\begin{align}
    \left(\frac{m}{3}\right)^3=\frac{32\pi^2}{g_{\rm YM}^2}
\end{align} 
thus the truncation is called as the BMN truncation.

\subsection{Symmetries and Classical Ground States}\label{sym class}

Why does the KK reduction of $\mathcal{N}=4$ SYM on $S^3 \times \mathbb{R}$ produce the BMN matrix model?  
From a physics perspective, we do not fully understand why this reduction works as it does. However, it seems that the symmetries of the mother theory and truncated theory strongly constrain the result. 
Recall that $\mathcal{N}=4$ SYM has $PSU(2,2|4)$ superconformal symmetry. The BMN truncation consists of restricting to the $SU(2)_R$ singlet sector by retaining only the lowest Kaluza--Klein modes on $S^3$. The remaining symmetry is $SU(2|4)$, which matches with the BMN matrix model having the same symmetry.
\footnote{When one keeps only the lowest spherical harmonics, the curvature of $S^3$ generates universal quadratic mass terms, while the underlying $SU(2)$ structure constants naturally produce the cubic couplings characteristic of the BMN deformation.
In this sense, the reduced theory is the unique maximally supersymmetric one-dimensional truncation compatible with the isometries of the sphere. This is analogous to the familiar reduction of $\mathcal{N}=4$ SYM on $T^3$, where keeping only the zero mode yields a matrix quantum mechanics resembling BFSS. In that case, however, the connection to D0-branes becomes precise only after performing three T-dualities and taking an appropriate decoupling limit, under which a wrapped D3-brane is reinterpreted as a D0-brane probing the dual torus with radii $R_i'=\alpha'/R_i$ (see, e.g. \cite{Blair:2023noj,Blair:2024aqz} and references therein). As far as we know, the sphere case has no direct analogue of such T-dualities.}

Specifically, the matrix model contains 16 dynamical and 16 kinematical supercharges. The dynamical supercharges form an $SU(2|4)$ superalgebra and transform nontrivially under the bosonic symmetry $SO(3)\times SO(6).$
The algebra of dynamical supercharges takes the form (see, e.g., \cite{Dasgupta:2002hx})
\begin{align}
\{ Q^{\dagger I \alpha}, Q_{J \beta} \} 
= 2\,\delta^{I}_{J}\delta^{\alpha}_{\beta} H 
- \frac{\mu}{3}\,\epsilon^{ijk}(\sigma^{k})^{\alpha}{}_{\beta}\,\delta^{I}_{J}\, M^{ij}
- \frac{i\mu}{6}\,\delta^{I}_{J} (\sigma^{ab})^{\alpha}{}_{\beta}\, M^{ab},
\end{align}
which corresponds precisely to the part of the $\mathcal{N}=4$ superconformal algebra whose supercharges are $SU(2)_R$ neutral.\footnote{Here $I=1,\dots,4$ labels an internal spinor index and should not be confused with the spacetime index.}

The additional 16 kinematical supercharges, denoted $q_{I\alpha}$ and $q^{\dagger I\alpha}$, satisfy
\begin{align}
\{ q^{\dagger I \alpha}, q_{J\beta} \} \;=\; \delta^I_J\,\delta^{\alpha}_{\beta}\,P^+,
\end{align}
and correspond to polarization-changing symmetries in the $U(1)$ center-of-mass sector, with
\[
q_{\alpha}^I \;=\; \Tr(\theta^I_{\alpha})\,.
\]
Since these supercharges act only on the decoupled center-of-mass degrees of freedom, they are often regarded as trivial and are therefore frequently neglected. As a result, the literature commonly focuses on the dynamical supercharges and refers to the BMN matrix model as having 16 supercharges.

It is important to note that the BMN truncation is a \emph{classically consistent} truncation of the full nonlinear equations of motion, but not a strict quantum truncation—higher-loop effects can reintroduce couplings to the discarded modes \cite{Kim:2003rza}. Furthermore, while the parent $\mathcal{N}=4$ SYM-theory enjoys full $U(N)$ gauge invariance, the reduced model admits spontaneous symmetry breaking. While the trivial vacuum preserves $U(N)$, more general classical configurations lead to symmetry breaking of the form
\[
U(N) \;\longrightarrow\; U(N_1)^{k_1} \times \cdots \times U(N_n)^{k_n}.
\]

The classical vacua are determined by
\begin{align}\label{class vacua}
X_i = 0, 
\qquad 
X_a = \frac{m}{3}\, J_a,
\end{align}
which minimize the potential term in the Lagrangian written as
\begin{align}
\frac{1}{2}\,\Tr\!\left[
\left(\frac{m}{3}X_a + i\,\epsilon_{abc}X_b X_c\right)^2
+ \frac{1}{2}\left( i [X_i,X_j] \right)^2
+ \left( [X_i,X_a] \right)^2
+ \frac{m}{6}\,(X_i)^2
\right].
\end{align}
The matrices $J_a$ are generators of an $SU(2)$ subalgebra satisfying $[J_a,J_b]=i\,\epsilon_{abc} J_c$. Therefore, each vacuum corresponds to a direct sum of $SU(2)$ representations of total dimension $N$, satisfying
\[
N \;=\; \sum_{i=1}^{K} k_i\,N_i,
\]
where $k_i$ denotes the dimension of each irreducible $SU(2)$ representation. In the M-theory dual description, these vacua correspond to fuzzy two-sphere configurations, representing multiple M2-branes expanded in the pp-wave background.
Note that the trivial vacuum corresponds to $K=1$, $k_1=1$ and $N_1=N$. (In other words, the generators are simply zero, $X_a=0$.)

\subsection{Supersymmetric black objects in BMN}\label{susy bh}

In this subsection, we briefly digress to discuss supersymmetric black holes (or black objects) in the BMN matrix model.
Although an explicit supersymmetric black hole metric in the BMN background is not currently known, the BMN matrix model and its Witten index nevertheless provide nontrivial information about how the entropy grows as a function of conserved charges. For previous studies, see \cite{Choi:2023vdm,Chang:2024lkw,Gadde:2025yoa}.

For simplicity, we focus on the trivial vacuum sector, in which the $SU(N)$ gauge symmetry remains unbroken.
We consider excitations around this vacuum and analyze the corresponding large-$N$ behavior of the Witten index.
A similar analysis can be done for other vacuum representations.
We will see that there exists a deconfined phase whose entropy scales as $O(N^2)$ and is captured by the Witten index.\footnote{In general, other representations are also expected to yield $O(N^2)$ entropy in a deconfined phase. Since the number of vacua is given by the number of partitions of $N$, which scales as $e^{O(N)}$, the total entropy in the deconfined phase will typically be $O(N^2)$ unless there are fine-tuned cancellations among the indices of different vacua.}

We choose $Q=Q^4_{-}$. BPS states of interests are annihialted by $Q$ and $Q^{\dagger}$, saturating the BPS bound.
\begin{align}
\{ Q, Q^{\dagger} \} 
= 2H 
- \frac{2\mu}{3}M^{12}
- \frac{\mu}{3}(M^{45}+M^{67}+M^{89}),
\end{align}
The Witten index is defined as
\begin{align}
    I=\Tr (-1)^F e^{-\beta\{Q,Q^{\dagger}\}}e^{-\Delta_1(Q_1+J)}e^{-\Delta_2(Q_2+J)}e^{-\Delta_3(Q_3+J)}
\end{align}
where $Q_{1,2,3}:=M^{45,67,89}$ and $J:=M^{12}$.
An integral representation of the Witten index in this sector can be written as
\begin{align}\label{eq: Witten index}
    I = \frac{\kappa^N}{N!}\oint \prod_{i=1}^N \frac{dz_i}{2\pi i z_i} 
    \prod_{j\neq i}^N 
    \frac{\big(z_i - z_j\big)\big(z_i - ab z_j\big)\big(z_i - bc z_j\big)\big(z_i - ca z_j\big)}
    {\big(z_i - abcz_j\big)\big(z_i - a z_j\big)\big(z_i - b z_j\big)\big(z_i - c z_j\big)} .
\end{align}
Here $a$, $b$, and $c$ denote the fugacities $e^{-\Delta_1}$, $e^{-\Delta_2}$, and $e^{-\Delta_3}$, respectively. For simplicity, we restrict to the equal-fugacity case
\[
a=b=c=e^{-\beta},
\]
which corresponds to the equal-charge sector.

The authors of \cite{Choi:2023vdm} showed that, in the limit $\beta\to 0$, there exists a saddle-point solution describing a \emph{small black hole phase}.
By ``small black hole'' we mean a regime in which the charges are parametrically smaller than $N^2$, but still scale in $O(N^2)$. Formally,
\[
N^{-\epsilon}\ll\frac{Q_1+J_L}{N^2} \ll 1,
\quad \text{for any } \epsilon>0 .
\]
This small black hole saddle can be obtained by solving Eq.~(5.4) of \cite{Choi:2021lbk}, which may be written equivalently as
\begin{align}
    (-1)^m \frac{1-f_m}{m}\,\rho_m = \text{const.}
\end{align}
Here $
1-f_m = (1-a^m)^3 \simeq m^3 \beta^3$ where $f_m$ coincides with $a_m$ in \cite{Choi:2021lbk}, and $\rho_m$ denotes the Fourier coefficients of the eigenvalue density,
\begin{align}
    \rho(\theta)
    &= \frac{1}{2\pi} + \frac{1}{\pi}\sum_{n=1}^{\infty} \rho_n \cos(n\theta).
\end{align}
Solving this equation yields
\[
\rho_m \propto (-1)^m \frac{1}{m^2}.
\]
which in turn leads to the eigenvalue distribution
\begin{align}
    \rho(\theta)
    &= \frac{3}{4\pi^3}(\pi^2-\theta^2),
    \qquad \theta\in(-\pi,\pi).
\end{align}

The large-$N$ free energy can then be computed as
\begin{align}
\log I
&= N^2 \sum_{n=1}^{\infty} \frac{1}{n}
\int_{-\pi}^{\pi} d\theta_1\, d\theta_2 \,
\bigl[f_n-1\bigr] e^{in(\theta_1-\theta_2)}\rho(\theta_1)\rho(\theta_2) \\
&= N^2 \sum_{n=1}^{\infty} \frac{f_n-1}{n}\,(\rho_n)^2
= -\frac{3N^2\beta^3}{2\pi^2}.
\end{align}
Performing a Legendre transformation, we obtain the entropy
\begin{align}
    S
    &= \underset{\beta}{\mathrm{Ext}}
    \left[
        -\frac{3N^2\beta^3}{2\pi^2}
        + \beta (Q+3J_L)
    \right] = \frac{2\pi}{9}\sqrt{\frac{2 (Q+3J_L)^3}{N^2}},
\end{align}
where $Q=Q_1+Q_2+Q_3$, and the saddle-point condition gives
\[
Q+3J_L = \frac{9N^2\beta^2}{2\pi^2}.
\]
Since $\beta$ is small in this regime, we indeed have $(Q+3J_L)/N^2 \ll 1$, confirming that this saddle corresponds to a small black hole phase.
Qualitatively, this behavior is analogous to small black holes in AdS$_5$.
In the canonical ensemble, this saddle is subleading due to its negative free energy, and it is accompanied by features such as negative specific heat \cite{Chang:2024lkw}.

On the other hand, the regime of finite charges in units of $N^2$ has been explored in \cite{Chang:2024lkw}, specifically where $(Q+3J_L)/N^2$ is fixed in the large $N$ limit. Numerical analysis confirms that in this regime, the entropy scales as $N^2$, following the form $S \sim N^2 f(\frac{Q+3J_L}{N^2})$. Namely, for a fixed charge in units of $N^2$, the entropy in units of $N^2$ is given by the constant.

Here, we demonstrate that the black-hole phase in $Q+3J_L\gg N^2$ in the BMN matrix model exhibits qualitative differences from its counterpart in four-dimensional holographic theories like \( \mathcal{N} = 4 \) SYM. This distinction is expected, since the single letter index of the BMN model does not diverge for any value of the fugacities. In contrast, for superconformal indices in $\mathcal{N}=4$ SYM, the large black hole phase is associated with fugacity regimes in which the single letter index diverges. This regime is often referred to as the Cardy limit \cite{Choi:2018hmj}; the BMN matrix model does not admit such a limit.

To probe the large-charge behavior, let us study the index \eqref{eq: Witten index}. Rather than solving the saddle-point equations directly—an approach we were unable to pursue—we estimate an upper bound on the index by computing the free BPS partition function.
\begin{align}
I < S_{\rm free~BPS}= \frac{\kappa^N}{N!}\oint \prod_{i=1}^N \frac{dz_i}{2\pi i z_i} \prod_{j\neq i}^N \frac{\big(z_i - z_j\big) \big(z_i + ab z_j\big) \big(z_i +bc z_j\big) \big(z_i +ca z_j\big)}{\big(z_i - abcz_j\big) \big(z_i - a z_j\big) \big(z_i - b z_j\big) \big(z_i - c z_j\big)} .
\end{align}
We take $a=b=c=e^{-\beta}$ with $\beta=0^{+}$.

Since the absolute values of $z_i+a^2 z_j$ are bounded from above by $2$, the free BPS partition function admits the bound
\begin{align}\label{hi}
S_{\rm free~BPS}< 2^{3N^2}\cdot \frac{\kappa^N}{N!}\oint \prod_{i=1}^N \frac{dz_i}{2\pi i z_i} \prod_{j\neq i}^N \frac{\big(z_i - z_j\big)}{\big(z_i - az_j\big)^4} .
\end{align}
At leading order, this is equivalent to $2^{3N^2}$ times the free partition function of four bosonic $U(N)$ matrices, all carrying the same fugacity $a$.

Matrix integrals of this type are known to admit a Hironaka decomposition, which expresses the generating function of gauge-invariant operators as a rational function whose denominator captures the algebraically independent generators (primaries), while the numerator accounts for a finite set of additional generators (secondaries) \footnote{See \cite{deMelloKoch:2025ngs} for a more detailed explanation.}. Concretely, the right-hand side of \eqref{hi} can be written in the form
\begin{align}
2^{3N^2}\cdot\frac{num}{den}
=2^{3N^2}\cdot \frac{1+\sum_i c_i a^i}{\prod_j(1-a^{j})^{n_{j}}},
\end{align}
where the denominator corresponds to the primary invariants and the numerator to the secondary invariants, with $c_i$ positive integers.

More precisely, for a system of $d$ bosonic $U(N)$ matrices, the invariant ring is finitely generated, and the number of algebraically independent primary generators is known to be
\[
\sum_{j}n_j=1+(d-1)N^2 .
\]
In our case, with $d=4$, this gives $1+3N^2$ primaries. The denominator therefore controls the asymptotic growth of the free partition function at large charge.
The number of secondary generators is finite but large: it is known to scale as $e^{cN^2}$ for some order-one constant $c$. 

At leading order, we therefore obtain the upper bound
\begin{align}
I<Z_{\rm free~BPS}<-3N^2\log \beta+(c+3\log2)N^2.
\end{align}
which gives the entropy bound as
\begin{align}
    S_{\rm ind}<S_{\rm free~BPS}<3N^2\log \left(\frac{Q+3J_L}{3N^2}\right)+(c+3\log2)N^2.
\end{align}
Since this is only an upper bound, the actual growth of the index may be slower. We thus expect that, at asymptotically large charge, the index behaves as
\begin{align}
\log I \sim -c_0 N^{\alpha}\log\beta +c_1N^2,
\end{align}
with $\alpha\leq 2$. The logarithmic dependence arises because the BMN index itself can be written as a rational function analogous to the free BPS partition function,
\begin{align}
I= \frac{num}{den},
\end{align}
where both the numerator and the denominator are finite polynomials in the fugacities. A difference from the free bosonic partition function is that the numerator now contains negative coefficients, reflecting fermionic contributions and cancellations in the index.
For instance, the $SU(3)$ BMN index at $a=b=c=t^2$ is given by \cite{Gadde:2025yoa}\footnote{To be precise, it is the $U(N)$ theory that admits a holographic interpretation. We nevertheless comment on the $SU(N)$ case below, since the large-$N$ entropy agrees at leading order. Moreover, the $U(1)$ contribution to the index is straightforward to compute and does not affect the leading behavior.}
\begin{align}
    I_{SU(3)} = & \ (1\!+\!5 t^2\!+\!24 t^4\!+\!82 t^6\!+\!245 t^8\!+\!628 t^{10}\!+\!1444 t^{12}\!+\!3013 t^{14}\!+\!5806t^{16}\!+\!10431 t^{18} \nonumber \\
&\!+\!17628 t^{20}\!+\!28174 t^{22}\!+\!42820 t^{24}\!+\!62104 t^{26}\!+\!86268 t^{28}\!+\!115035 t^{30}\!+\!147595 t^{32}\nonumber \\
&\!+\!182506 t^{34}\!+\!217831 t^{36}\!+\!251223 t^{38}\!+\!280233 t^{40}\!+\!302516 t^{42}\!+\!316224t^{44}\!+\!320131 t^{46}\nonumber \\
&\!+\!313920 t^{48}\!+\!298116 t^{50}\!+\!274103 t^{52}\!+\!243875t^{54}\!+\!209819 t^{56}\!+\!174384 t^{58}\!+\!139852 t^{60} \nonumber \\
&\!+\!108049 t^{62}\!+\!80282t^{64}\!+\!57228 t^{66}\!+\!39035 t^{68}\!+\!25393 t^{70}\!+\!15687 t^{72}\!+\!9153 t^{74}\!+\!5013 t^{76} \nonumber \\
&\!+\!2553 t^{78}\!+\!1197 t^{80}\!+\!507 t^{82}\!+\!189 t^{84}\!+\!60 t^{86}\!+\!15 t^{88}\!+\!3 t^{90} ) \nonumber \\
&\times \frac{(1-t^2)^5 (1-t^4)^3 }{(1-t^8)^3 (1-t^{10})^2 (1-t^{12}) (1-t^{14})^3 (1-t^{18})}.
\end{align}
At very large charges, beyond the maximal degree of the numerator, the contribution of the numerator to the partition function saturates. The denominator, however, generates an infinite series, and the poles of the index give rise to logarithmic divergences.
For $SU(3)$, the fastest divergence occurs as $t$ approaches $t\to e^{\pi i/4}$, for which $\log I\sim 3\log (Q+3J_L)$

The (indicial) entropy is obtained by a Legendre transform,
\begin{align}
S_{\rm ind}
\sim \underset{\beta}{\mathrm{Ext}}
\left[
\log Z
+ \beta (Q+3J_L)
\right]
\sim c_0 N^\alpha \log\frac{Q+3J_L}{N^{\alpha}} + c_1 N^2 .
\end{align}
For large $N$ the first term dominates when $\alpha=2$, whereas the second term dominates for $\alpha<2$.

From the gravitational perspective, we expect $\alpha<2$. If $\alpha=2$, this would suggest the existence of a black hole whose horizon area grows logarithmically with the charges. Such behavior is not expected in general for black objects, whose horizon areas typically scale polynomially with the charges. 

Let us note the logarithmic growth of $\log I$ for the gauge groups $SU(2)$, $SU(3)$, $SU(4)$, $SU(5)$, and $SU(6)$ is given by $3\log j$, $2\log j$, $8\log j$, $6\log j$, and $16\log j$, respectively, where $j\equiv Q+3J_L$. From these data alone, it is difficult to determine whether $\alpha$ saturates the bound $\alpha=2$. It would be interesting to determine the value of $\alpha$ at large $N$, either numerically or analytically, which we leave for future work.

We therefore expect that any consistent bulk interpretation corresponds to a configuration in which the horizon area ceases to grow beyond some value of the charge. In other words, at fixed $N$, the radius of a BPS black hole in the pp-wave background is bounded. For larger charges, the geometry is expected to transition into a charged, hairy black hole configuration. 
Therefore, to obtain higher entropy at charge $Q$, one must take a large-$N$ limit in which
\[
N \gtrsim O(Q^{1/2}).
\]
The precise nature of the resulting hairy black hole remains unclear, and it would be interesting to investigate its structure and properties in future work.

In short, black objects in matrix models are very different from the BPS black holes in AdS, which can grow indefinitely. Heuristically, we believe this is because plane-wave backgrounds impose a rigid harmonic potential, which suppresses the number of excitations within a given energy window and therefore limits the available entropy, resulting in a smaller horizon.

\section{BMN-like matrix models}\label{bmnlike}

In this section, we propose a class of metric which is a generalization of the pp-wave geometry \eqref{ppmetric}.
Then, we propose a class of matrix models and conjecture that they are dual to the DLCQ of $M$ theory in this background.

As explained in the introduction briefly, we will be interested in the following of pp-wave--like metric, where we replace the metric of the $S^5$ in \eqref{sphere ppwave} to that of a five dimensional space $Y_5$, written as
\begin{align}\label{ppwave2}
    ds^2=-2dx^{+}dx^{-} +\sum_{a=1}^{3}(dx^a)^2+dr^2+r^2dY_5^2-\left[\sum_{a=1}^{3}\frac{\mu^2}{9}x_a^2+\frac{\mu^2}{36}r^2\right](dx^+)^2,
\end{align}
where $dY_5^2$ refers to a metric of $Y_5$. We consider $Y_5$ to be Sasaki-Einstein or their cousins.
The flux $F_{+123}=\mu$ remains unchanged.

For the matrix theories, we consider BMN truncations of $4d$ superconformal theories dual to type IIB string theory on AdS$_5\times Y_5$. Let us note again that this truncation is a classically consistent truncation.

\subsection{Generalized pp-wave}

First, we confirm that the metric \eqref{ppwave2} satisfies the Einstein-Maxwell equation written as
\begin{align}
R_{MN} = \frac{1}{12}\left(F_{MPQR}F_N{}^{PQR} - \frac{1}{12}g_{MN} F^2\right), 
\quad dF = 0, 
\quad d*F + \frac{1}{2} F\wedge F = 0.
\end{align}
From the metric, the only nontrivial curvature component is $R_{++}$, arising from the potential term in $H$, such that $R_{++}=\frac{1}{2}\nabla^2H$. 
The left-hand side of the Einstein equation is therefore
\begin{align}
R_{++}
= \frac{1}{2}\nabla^2 H 
= \frac{1}{2}\left(\sum_{a=1}^{3}\partial_a^2 H + \nabla^2_{CY_3} H\right)
= \frac{1}{2}\left(3\cdot \frac{2\mu^2}{9} + 6\cdot\frac{2\mu^2}{36}\right) 
= \frac{\mu^2}{2},
\end{align}
since $H$ depends only on $x_a^2$ and $r^2$.
The right-hand side of Einstein’s equation is
\begin{align}
\frac{1}{12} F_{+IJK} F_{+}{}^{IJK} 
= \frac{1}{12} \mu^2 \cdot 3! 
= \frac{\mu^2}{2},
\end{align}
which agrees with the left-hand side, confirming that the metric satisfies the Einstein equation.

Next, we confirm that the metric preserves supersymmetry.  
In ``spherical'' coordinates, the Killing spinor equation takes the same form as in the pp-wave background, as written in \eqref{kispin} and \eqref{kispin2}, except that the metric for the angular directions $m$ is different. 
One is therefore led to solve
\begin{align}\label{sasa}
    \tilde{\nabla}_m S
    - \frac{1}{2r}\,\Gamma_{mr}\, S = 0,
\end{align}
where $\tilde{\nabla}_m$ denotes the intrinsic covariant derivative on $\mathrm{SE}_5$.

On a Sasaki--Einstein five--manifold, Killing spinors satisfying \eqref{sasa} are known to exist, with the number of solutions depending on the geometry.  
For the round $S^5$, one obtains the maximal set of Killing spinors, whereas for a generic toric Sasaki–Einstein manifold $M_5$ this number is reduced by a factor of $1/4$ compared to $S^5$. Consequently, the generalized pp-wave arising from \eqref{ppwave2} typically preserves $8$ of the original $32$ supercharges, comprising $4$ dynamical and $4$ kinematical supercharges.

\subsubsection{Penrose limit from AdS$_4\times X_7$}

Here we show that the generalized pp-wave \eqref{ppwave2} can be obtained from a Penrose limit of AdS$_4\times X_7$, where $X_7$ is an Einstein manifold satisfying
\begin{equation}
\operatorname{Ric}_{mn}(X_7)
    = \Lambda_7\, g_{mn}.
\end{equation}
We use the Freund--Rubin ansatz
\begin{equation}
ds^2_{11}
    = ds^2_{\mathrm{AdS}_4(L)} + ds^2_{X_7},
    \qquad
F_{(4)} = f\, \mathrm{vol}(\mathrm{AdS}_4),
\end{equation}
which solves the eleven--dimensional supergravity equations provided that $f^2 = \frac{9}{L^2}$ and
the metric for $X_7$ is written as 
\begin{equation}\label{eq:X7metric}
ds^2_{X_7}
    \;=\;
    (2L)^2\left(\cos^2\!\theta\, d\psi^2
    + d\theta^2
    + \sin^2\!\theta\, ds^2_{M_5}\right),
    \qquad
    0 < \theta < \pi,
\end{equation}
where $ds^2_{M_5}$ is a fixed metric on a five--dimensional manifold $M_5$.
We choose $M_5$ to be SE$_5$, normalized so that its Ricci tensor is
$\operatorname{Ric}_{g_5}=4 g_5$ (the usual Sasaki--Einstein convention).  
One can explicitly verify that the ansatz \eqref{eq:X7metric} produces a local Einstein seven--metric with
\begin{equation}
\operatorname{Ric}(X_7)=\frac{6}{(2L)^2}\, g_{X_7},
\end{equation}
which is precisely the condition $\Lambda_7 = \frac{3}{2L^2}$ required by the Freund--Rubin equations.
Taking the Penrose limit around $\theta=0$ reduces the geometry to the pp-wave--like metric \eqref{ppwave2}.
Let us note that \eqref{eq:X7metric} reproduces the round $S^7$ metric if $M_5$ is $S^5$.

Let us construct the cone over $X_7$, omitting the overall $2L$ scale factor for simplicity:
\begin{align}
    ds^2_{C(X_7)} = dr^2 + r^2 ds^2_{X_7} = dr^2 + r^2(d\theta^2 + \sin^2\!\theta\, ds^2_{M_5} + \cos^2\!\theta\, d\psi^2)
\end{align}
By introducing the coordinate transformations $r_1 = r \sin\theta$ and $r_2 = r \cos\theta$, we can neatly rewrite the cone metric as a direct product:
\begin{align}
    ds^2_{C(X_7)} = (dr_1^2 + r_1^2 ds^2_{M_5}) + (dr_2^2 + r_2^2 d\psi^2),
\end{align}
which shows that the cone over $X_7$ is the Cartesian product of the cone over $M_5$ and $\mathbb{R}^2$.
Therefore, one can obtain the AdS$_4\times X_7$ geometry by putting M2-branes at the apex of a transverse 8-dimensional space that is the Cartesian product of a 6D Calabi–Yau manifold and a flat 2D plane. This must be dual to a 3d $\mathcal{N}=2$ matter-Chern-Simons theory, which we leave for a future study to write down the explicit Lagrangian.

\subsection{BMN truncation from holographic gauge theories}\label{holo}

In this subsection, we study the BMN truncation of holographic (quiver) gauge field theories \cite{Klebanov:1998hh,Martelli:2004wu,Benvenuti:2004dy,Kachru:1998ys,Lawrence:1998ja}, focusing on $Y^{p,q}$ and $\mathbb{Z}_p$ orbifold theories, and $\beta$-deformed $\mathcal{N}=4$ SYM.

In the first subsubsection, we study the $Y^{p,q}$ case and show that the global symmetries of the resulting matrix quantum mechanics agree with those of the corresponding eleven-dimensional dual backgrounds.

In the second subsubsection, we turn to orbifold theories, where the structure is more transparent. We explain how this construction can be viewed as arising from orbifolding the type IIA background that supports D0-branes.

Finally, we discuss a theory that arises from marginal deformations of $\mathcal{N}=4$ SYM.

\subsubsection{$Y^{p,q}$}
The \(Y^{p,q}\) theories are four-dimensional \(\mathcal{N}=1\) quiver gauge theories that arise as the worldvolume theories of D3-branes placed at the tip of the Calabi--Yau cone over the five-dimensional Sasaki--Einstein manifold \(Y^{p,q}\) \cite{Martelli:2004wu,Benvenuti:2004dy}. Holographically, these theories are dual to type IIB string theory on \(\mathrm{AdS}_5\times Y^{p,q}\). The quiver for a general \(Y^{p,q}\) model contains \(2p\) gauge nodes (each an \(SU(N)\) factor) and a collection of bifundamental chiral multiplets arranged in a periodic pattern whose detailed connectivity depends on \((p,q)\). The theory has \(4p + 2q\) bifundamental chiral multiplets of four types:
\(p\) fields \(U_\alpha^{(k)}\) (\(\alpha=1,2\)) of type \(U_{1,2}\), \(q\) fields \(V_\alpha^{(k)}\) (\(\alpha=1,2\)) of type \(V_{1,2}\),  \(p - q\) fields \(Z^{(k)}\) of type \(Z\), and \(p + q\) fields \(Y^{(k)}\) of type \(Y\).
All matter fields are bifundamentals connecting adjacent \(SU(N)\) gauge nodes according to the quiver pattern.
The superpotential is obtained by summing all cubic and quartic gauge–invariant loops involving \(U_\alpha^{(k)}, V_\alpha^{(k)}, Y^{(k)}, Z^{(k)}\).
Explicitly \cite{Benvenuti:2004dy},
\begin{align}
W
&= \sum_{k} \epsilon^{\alpha\beta}\,
\Tr\!\Big(
    U_{\alpha}^{(k)} V_{\beta}^{(k)} Y^{(k+2)}
  + V_{\alpha}^{(k)} U_{\beta}^{(k+1)} Y^{(2k+3)}
\Big)
+ \sum_{k} \epsilon^{\alpha\beta}\,
\Tr\!\Big(
    Z^{(k)} U_{\alpha}^{(k+1)} Y^{(2k+3)} U_{\beta}^{(k)}
\Big),
\label{Ypq_superpotential}
\end{align}
where the antisymmetric tensor \(\epsilon^{\alpha\beta}\) contracts the \(SU(2)\) flavor indices of \(U_\alpha, V_\alpha\).
Each trace corresponds to a closed plaquette in the quiver diagram.
When $p=1, q=0$ it reduces to the Klebanov-Witten theory $T^{1,1}$ \cite{Klebanov:1998hh}.

\vspace{0.5em}

Let us show, following \cite{Asplund:2015yda}, that the BMN truncation yields a classically consistent reduction for these $\mathcal{N}=1$ theories. \footnote{The paper \cite{Asplund:2015yda} have studied the BMN truncation of various $4d$ $\mathcal{N}=1$ supersymmetric quiver gauge theories, although they have not studied its M-theory duals. The paper focuses on massive quiver matrix models that describe the dynamics of non-relativistic charged particles in global AdS$_4$.}
The key point is identical to the \(\mathcal{N}=4\) case: after inserting the truncation ansatz every term appearing in each equation of motion has the same harmonic (coordinate) dependence on \(S^3\), so the angular dependence factors out and the truncated modes form a closed set. 

To illustrate this, consider the scalar equation of motion (for a generic chiral multiplet \(\phi^i\)):
\begin{align}\label{n=1 scalar}
    D^\mu D_\mu \bar{\phi}^i \;+\; \frac{\partial V}{\partial \phi^i}
    \;+\; \sqrt{2}\, g\, (\lambda\,\bar{\psi}^i)
    \;+\; \frac{1}{2}\, \frac{\partial^3 W}{\partial \phi^i \partial \phi^j \partial \phi^k}\, \psi^j \psi^k \;=\; 0,
\end{align}
where \(V\) denotes the scalar potential (including both \(F\)- and \(D\)-term contributions), \(\lambda\) the gaugino, and \(\psi^i\) the chiral fermions. We apply the BMN truncation ansatz exactly as in \eqref{ansatz}, i.e. retain only the spatially constant (\(k=0\)) scalar harmonics, the lowest spinor harmonics for fermions, and the three Killing vectors/Killing spinors for the gauge field components on \(S^3\) that are $SU(2)_R$ singlet.

A few remarks that justify the consistency of the reduction in \eqref{n=1 scalar}:

\begin{itemize}
  \item The covariant Laplacian term \(D^\mu D_\mu \bar{\phi}^i\) evaluated on the ansatz is independent of the angular coordinates because the vector harmonic normalization implies \(V^{\hat{a}a}V_{a}^{\hat{b}}=\delta^{\hat{a}\hat{b}}\). Thus this contribution is a sphere-constant function multiplying the time-dependent matrix degree of freedom.
  \item The potential derivative \(\partial V/\partial\phi^i\) is constructed from (anti)commutators and products of scalars that are taken to be \(k=0\) modes; therefore it is manifestly constant on \(S^3\).
  \item Fermion bilinears such as \((\lambda \bar{\psi}^i)\) reduce consistently because the spinor harmonic orthonormality gives \(S^{\hat{\alpha}\dagger}S^{\hat{\beta}}=\delta^{\hat\alpha\hat\beta}\), producing the same harmonic structure for each such term.
  \item Yukawa-type terms coming from third derivatives of \(W\) reduce properly because spinor harmonic identities like \(S^{\hat{\alpha}\top} i\sigma^2 S^{\hat{\beta}} = i(\sigma^2)^{\hat{\alpha}\hat{\beta}}\) ensure that the angular dependence factors into known constant tensors. Useful identities of this form are collected in Appendix B of \cite{Kim:2003rza}.
\end{itemize}

\vspace{0.5em}

The gauge-multiplet sector of the reduced quantum-mechanical action is universal: for each gauge node one obtains the same bosonic and fermionic kinetic terms, mass terms coming from the curvature coupling, the Myers-type cubic term sourced by the \(S^3\) Killing vector algebra, and the usual quartic commutator potential. If the quiver contains multiple gauge nodes the full action is the sum of the contributions for each node, with bifundamental matter coupling the different node sectors according to the quiver connectivity. The gauge-multiplet part of the one-dimensional Lagrangian (per gauge node) takes the expected BMN form:
\begin{align}\label{gauge}
L_{\rm gauge}=\text{Tr} \left( \frac{1}{2} (D_t X^a_v)^2 + \frac{1}{4} [X^a_v, X^b_v]^2 + \frac{i}{2} (\lambda^\dagger_v D_t \lambda_v - (D_t \lambda^\dagger_v)\lambda_v) - \lambda^\dagger_v \sigma^i [X^a_v, \lambda_v] \right)\nonumber\\
- \text{Tr} \left( \frac{1}{2} \left(\frac{m}{3}\right)^2 (X^a_v)^2 + \frac{m}{2} \lambda^\dagger_v \lambda_v + i \frac{m}{3} \epsilon_{abc} X^a_v X^b_v X^c_v \right)
\end{align}
where $a,b,c\in (1,2,3)$ and $v$ labels a gauge node.

On the other hand, bifundamental chiral multiplets reduce to matrix-valued complex scalars and their fermionic partners; their kinetic terms, mass terms (from the curvature coupling), and Yukawa interactions descend from the parent superpotential and follow the same consistent-harmonic pattern described above. 
Summing over all nodes and including the reduced matter sector yields a multi-matrix quantum mechanics whose interactions are fixed by the quiver superpotential and whose bosonic sector contains the characteristic BMN mass and Myers couplings.

While we dimensionally reduce the UV Lagrangian to quantum mechanics, we retain the R-charge assignments associated with the 4d IR fixed point. This choice, similar to the dimer construction in \cite{Franco:2005rj}, ensures that the R-symmetry of the 1d theory correctly encodes the isometries of the underlying Sasaki-Einstein geometry.


To be specific, the part of the Lagrangian involving the bifundamental matter fields is given by (see, e.g., Section 2 of \cite{Asplund:2015yda}).
\begin{align}\label{matter}
L_{\rm matter} &= \sum_i \text{Tr} \left( |D_t \phi^i|^2 + |F^i|^2 + \frac{i}{2} (\psi^{i\dagger} D_t \psi^i - (D_t \psi^{i\dagger})\psi^i)+i\frac{m}{6} \left( (D_t \phi^{i\dagger})\phi^i - \phi^{i\dagger} D_t \phi^i \right) - \frac{m}{6}  \psi^{i\dagger} \psi^i  \right) \nonumber\\
 &+ \sum_i \text{Tr} \left( \frac{\partial W}{\partial \phi^i} F^i + h.c. \right) + \sum_{ij} \text{Tr} \left( \frac{1}{2} \frac{\partial^2 W}{\partial \phi^i \partial \phi^j} \psi^j \psi^i + h.c. \right) \nonumber\\
 & - \sum_{i:v\to w} \text{Tr} \Big( (\phi^{j\dagger} X^a_w - X^a_v \phi^{i\dagger})(X^a_w \phi^i - \phi^i X^a_v) - \phi^{i\dagger}(D_w \phi^i - \phi^i D_v) \nonumber \\
&\quad + \psi^{i\dagger} \sigma^a (X^a_w \psi^i - \psi^i X^a_v) + i\sqrt{2} \big( (\phi^{i\dagger} \lambda_w - \lambda_v \phi^{i\dagger}) \psi^i - \psi^{i\dagger} (\lambda^\dagger_w \phi^i - \phi^i \lambda^\dagger_v) \big) \Big)
\end{align}
Here $i$ labels a bifundamental multiplet, and the notation $i:v\to w$ indicates that $\phi^i$ transforms in the fundamental representation of the gauge group at node $v$ and in the anti-fundamental representation of the gauge group at node $w$.
Also, $D_t$ is defined as $D_t=\partial_t - i[A,\cdot] - i r m$, where $r$ denotes the $R$-charge.
This class of Lagrangians preserves four real supercharges and realizes an \(su(2|1)\) superalgebra \cite{Asplund:2015yda}, matching the four dynamical Killing spinors in the pp-wave–like background.

Because the gauge-sector Lagrangian is unchanged from that of $\mathcal{N}=4$ SYM (essentially the Lagrangian of $\mathcal{N}=1$ vector multiplets), the classical vacua are given by
\begin{align}\label{vacua 2}
\phi^i=0, \quad X^a_v\propto J^a_v,
\end{align}
where the index $v$ labels the gauge nodes. This coincides with the BMN vacua discussed in \eqref{class vacua}. This follows from the fact that, when all bifundamental fields are set to zero, the potential reduces to $\Tr\!\left[\left(\frac{m}{3}X_a + i\,\epsilon_{abc}X_b X_c\right)^2\right]$.
Finally, since there are $2p$ gauge nodes, the momentum along the light-cone direction scales as $P^{+}\propto N\cdot 2p$.

On the other hand, the explicit form of the reduced potential and Yukawa couplings depends on the detailed quiver data and on the superpotential combinatorics determined by \((p,q)\). Thus writing an explicit reduced Lagrangian for a specific \(Y^{p,q}\) model requires inserting the ansatz to the original action and integrate over the sphere.

\subsubsection{$\mathbb{Z}_p$ Orbifold of $\mathcal{N}=4$ SYM}
Here, we study the $\mathbb{Z}_p$ orbifold of $\mathcal{N}=4$ super Yang--Mills theory, which is dual to type~IIB string theory on $\mathrm{AdS}_5 \times S^5/\Gamma$. The orbifold group $\Gamma=\mathbb{Z}_p$ acts on the transverse complex coordinates
\[
(z_1,z_2,z_3) \equiv (x_4+i x_5,\; x_6+i x_7,\; x_8+i x_9)
\]
according to
\begin{align}\label{ident}
\Gamma:\ (z_1,z_2,z_3) \mapsto (z_1\,\omega^{a_1},\; z_2\,\omega^{a_2},\; z_3\,\omega^{a_3}),
\end{align}
where $\omega=e^{2\pi i/p}$ and supersymmetry requires 
\[
a_1+a_2+a_3 \equiv 0 \quad (\mathrm{mod}\; p).
\]
When all $a_i\not\equiv0$, the theory typically preserves $\mathcal{N}=1$ supersymmetry; if one $a_i$ vanishes (with the supersymmetry constraint still satisfied), the orbifold preserves $\mathcal{N}=2$.  
As a notable example, the $\mathbb{Z}_{2p}$ orbifold with $(a_1,a_2,a_3)=(1,1,-2)$ yields the $Y^{p,p}$ quiver gauge theory.

In the dictionary developed above, the BMN truncation of the orbifold gauge theory corresponds holographically to orbifolding the pp-wave geometry \eqref{ppmetric} by the same action \eqref{ident}. This identification becomes transparent from the D0-brane point of view.  

Start with $Np$ D0-branes propagating in the IIA background obtained by dimensional reduction of the eleven-dimensional plane-wave. The matrices $X_4,\dots,X_9$ encode the transverse coordinates of these D0-branes, and imposing the identification \eqref{ident} on these coordinates corresponds to quotienting the transverse $\mathbb{C}^3$ by the same $\mathbb{Z}_p$ action that defines the field-theory orbifold. In the matrix model, this projection acts simultaneously on the geometric degrees of freedom and on the Chan--Paton factors, reducing the parent $U(Np)$ theory to a quiver structure.

The parent $U(Np)$ matrix model contains three adjoint complex scalars
\[
(\Phi_1,\Phi_2,\Phi_3)\equiv(X_4+iX_5,\; X_6+iX_7,\; X_8+iX_9),
\]
each an $(Np)\!\times\!(Np)$ matrix.  
The orbifold acts both on the matrices and on the gauge indices.  
We embed $\Gamma$ in the gauge group through the Chan--Paton matrix
\begin{equation}
\gamma \;=\; 
\mathrm{diag}\!\left(
\mathbf{1}_N,\;
\omega\,\mathbf{1}_N,\;
\omega^2\,\mathbf{1}_N,\;
\dots,\;
\omega^{p-1}\mathbf{1}_N
\right),
\qquad \gamma^p=\mathbf{1}_{Np},
\end{equation}
which breaks
\begin{equation}
U(Np) \;\longrightarrow\; \prod_{i=0}^{p-1} U(N)_i.
\end{equation}

The geometric action on $\mathbb{C}^3$ is
\begin{equation}
\Phi_a \;\longrightarrow\; \omega^{a_a}\,\Phi_a,
\qquad a=1,2,3,
\end{equation}
subject to the supersymmetry constraint $a_1+a_2+a_3\equiv0\ (\mathrm{mod}\; p).$

A matrix $\Phi_a$ survives the orbifold projection if it is invariant under the combined geometric and Chan--Paton action:
\begin{equation}\label{proj-condition}
\Phi_a \;=\; \omega^{a_a}\,\gamma\,\Phi_a\,\gamma^{-1}.
\end{equation}
Writing $\Phi_a$ in $N\times N$ blocks, $
(\Phi_a)_{ij},~ i,j=0,\dots,p-1,$
the condition \eqref{proj-condition} becomes
\begin{align}
(\Phi_a)_{ij}\neq 0
\quad\Longleftrightarrow\quad
\omega^{a_a + i - j}=1
\quad\Longleftrightarrow\quad
j-i\equiv a_a\;(\mathrm{mod}\;p).
\end{align}
Thus each $\Phi_a$ decomposes into bifundamental components
\begin{equation}
\Phi_a:\quad (\Phi_a)_{i,i+a_a} \in (N_i,\overline{N}_{\,i+a_a}),
\end{equation}
with indices understood modulo $p$.  
In particular,
if $a_a = 0$, then $\Phi_a$ survives only on the diagonal and transforms in the adjoint of each $U(N)_i$.
If $a_a \neq 0$, then $\Phi_a$ becomes a bifundamental linking node $i$ to node $i+a_a$ in the quiver.

Since the orbifold action commutes with the KK reduction that defines the BMN truncation, one may orbifold first and reduce later, or vice versa—the result is the same. The combined projection produces a quiver matrix quantum mechanics with gauge group
\[
\prod_{i=0}^{p-1} U(N)_i,
\]
and with matter content determined entirely by the integers $(a_1,a_2,a_3)$.  
This is precisely the matrix-model counterpart of orbifolding the pp-wave geometry by $\Gamma$.

\subsubsection{$\beta$ deformed $\mathcal{N}=4$ SYM}

The $\beta$-deformation of $\mathcal{N}=4$ SYM is an exactly marginal deformation that preserves $\mathcal{N}=1$ supersymmetry while reducing the original $SU(4)_R$ symmetry to
\[
U(1)_R \times U(1) \times U(1).
\]
The exactly marginal nature of this deformation was first established in the original work of Leigh and Strassler \cite{Leigh:1995ep}. The deformed theory can be written in terms of three chiral superfields $\Phi_1$, $\Phi_2$, and $\Phi_3$ in the adjoint representation of the gauge group, with the superpotential
\begin{align}
W \;=\; h\,\Tr\!\left(
e^{i\pi\beta}\,\Phi_1\Phi_2\Phi_3
\;-\;
e^{-i\pi\beta}\,\Phi_1\Phi_3\Phi_2
\right),
\end{align}
where $\beta$ is a real deformation parameter and $h$ is related to the gauge coupling at the conformal fixed point.  
For real $\beta$, the theory remains conformal to all orders in perturbation theory and defines an exactly marginal deformation along the Leigh–Strassler fixed line.

On the gravity side, the dual background is obtained by applying a sequence of T-duality–shift–T-duality (TsT) transformations to a $U(1)\times U(1)$ torus inside the $S^5$ factor of the $\mathrm{AdS}_5 \times S^5$ geometry \cite{Lunin:2005jy}.  
This transformation acts by an $SL(2,\mathbb{R})$ rotation on the complexified Kähler modulus of the two-torus, corresponding geometrically to a shift and twist of two angular directions inside the five-sphere.

The resulting Lunin–Maldacena background preserves the same $\mathcal{N}=1$ supersymmetry and the same global
\[
U(1)_R \times U(1) \times U(1)
\]
isometries as the deformed gauge theory.  
The internal manifold is no longer Sasaki–Einstein, as the TsT transformation distorts the round $S^5$ metric and introduces nontrivial NS--NS and R--R two-form fluxes.  
The field-theory deformation parameter $\beta$ maps directly to these background flux components, encoding the twisted geometry of the deformed compactification.

One can apply an $SL(2, \mathbb{R})$ solution-generating ansatz to the maximally supersymmetric 11D pp-wave geometry:
\begin{align}
    ds^2=-2dx^{+}dx^{-} +\sum_{a=1}^{3}(dx^a)^2+dr^2+r^2d\Omega_{5}^2-\left[\sum_{a=1}^{3}\frac{\mu^2}{9}x_a^2+\frac{\mu^2}{36}r^2\right](dx^+)^2
\end{align}
The transverse space exhibits an $SO(6)$ symmetry, which contains the subgroup $U(1) \times U(1) \subset SU(3) \subset SO(6)$. 
By treating the torus formed by these two $U(1)$ isometries as an effective 9D compactification space, one can perform an $SL(2, \mathbb{R})$ transformation on its moduli to generate a new supergravity solution consisting of a deformed metric and non-trivial 3-form flux. Furthermore, because the original pp-wave possesses Killing spinors that are uncharged under this $U(1) \times U(1)$ action, the Lunin-Maldacena theorem guarantees that the deformed background will preserve the fraction of supersymmetry associated with those invariant supercharges.

In \cite{Shimada:2008xy}, the author establishes the microscopic dual to this geometry by demonstrating that this exact background emerges as the continuum limit of a $\beta$-deformed matrix model. Instead of applying the geometric transformation directly, the author derives the background fields by matching the deformed matrix Hamiltonian to the supermembrane action.

\subsection{D0 branes probing \(\mathbb{R}^{1,3}\times\mathrm{CY}_3\)}\label{subsec:D0_on_CY3}

Since the BFSS matrix model is obtained by taking the mass parameters of the BMN matrix model to zero, it is natural to consider analogous BFSS-like matrix quantum mechanics in more general backgrounds.

In particular, quiver matrix models arising as the zero–mass limits of BMN-like matrix models describe the low-energy dynamics of $N$ D0-branes in type IIA string theory probing
\[
\mathbb{R}^{1,3}\times \mathrm{CY}_3,
\]
where \(\mathrm{CY}_3\) is a non-compact Calabi–Yau cone whose base is a Sasaki–Einstein manifold.
In this sense, they provide a curved-space analogue of the BFSS matrix theory. 

Such quiver quantum mechanics have been extensively studied. Beginning with the work of Douglas and Moore \cite{Douglas:1996sw}, D-branes probing Calabi–Yau singularities were shown to give rise to quiver gauge theories, and their one-dimensional reductions describe the dynamics of D0-brane bound states (See \cite{Douglas:2000qw,Denef:2002ru,Aspinwall:2004jr,Franco:2005rj,Chuang:2009crq,Ooguri:2009ijd} and references therein).

Consider type IIA string theory on the ten-dimensional background
\(
\mathcal{M}_{10}=\mathbb{R}^{1,3}\times \mathrm{CY}_3.
\)
A stack of $N$ coincident D0-branes consists of pointlike objects extending only along the time direction; they are localized both in $\mathrm{CY}_3$ and in the spatial $\mathbb{R}^3\subset \mathbb{R}^{1,3}$. The worldline theory on the D0-brane stack is a one-dimensional supersymmetric quantum mechanics arising from open strings ending on the branes. When the D0-branes sit at a Calabi–Yau singularity, the light open-string modes reorganize into multiple species of fractional branes, and the low-energy dynamics is captured by a quiver quantum mechanics determined by the local singular geometry.

The Lagrangian of this quiver matrix quantum mechanics may be obtained from the general D0-brane worldvolume action by replacing the adjoint matrix coordinates with bifundamental fields and imposing the F-term interactions encoded by the superpotential of the singularity. This can be obtained by simply setting the mass terms in \eqref{gauge} and \eqref{matter} to zero.

The system can be generalized by including additional branes. Introducing D2-branes wrapping vanishing 2-cycles and D6-branes filling $\mathrm{CY}_3$ modifies the quiver and its Lagrangian, and the resulting theory describes bound states of D0/D2/D6 branes.

One can introduce supersymmetric mass deformations to these theories, following the general construction of \cite{Asplund:2015yda}. Such deformations play a role analogous to the BMN deformation of the BFSS model, lifting flat directions while preserving supersymmetry.

It would be interesting to identify the corresponding eleven-dimensional description and determine whether these mass-deformed quiver theories admit an interpretation in terms of an M-theory pp-wave–like background.

\subsection{Other theories}

One can dimensionally reduce a generic four-dimensional $\mathcal{N}=1$ supersymmetric gauge theory to a one-dimensional supersymmetric quantum mechanics. Here, we restrict attention to theories that flow to interacting conformal fixed points in the infrared, since such IR CFTs are natural candidates for admitting holographic descriptions. However, most of these theories do not allow low-energy supergravity descriptions, and the structure of a potential holographic dual remains unclear. 

Nevertheless, it remains useful to comment on certain general features of the resulting matrix models that may admit pp-wave–like supergravity duals and to extract lessons that could guide future investigations.
In this Subsection, we mainly study the truncated theories by examining their trivial vacuum and their Witten index.

\subsubsection{$\mathcal{N}=2$ SCFTs}

The first example we consider is the $\mathcal{N}=2$ $SU(N)$ gauge theory with $N_f = 2N$ hypermultiplets, which is a superconformal field theory even at the UV Lagrangian level.
In the large $N$ limit, the \emph{flavor-singlet sector} of this theory has been analyzed and claimed to be dual to a closed string theory, which does not possess a well-behaved supergravity limit.
The dual background is not of the simple product form AdS$_5 \times M_5$, but rather corresponds to a subcritical closed string theory in seven dimensions \cite{Gadde:2009dj}.
This is reflected in the Hagedorn-like growth of the superconformal index, indicating towers of infinitely many supersymmetric higher spin states \cite{Gadde:2009dj}.

The BMN truncation reduces the number BPS states drastically as the dimensional reduced theory does not have spatial derivatives that make a tower of infinite states. As shown below, there is no Hagedorn growth in the BMN index for this theory. We define index with two chemical potentials as 
\begin{align}
    \mathcal{I}=\Tr (-1)^F t^{2(\Delta+j)}v^{r-R}
\end{align}
where $\Delta$, $j$, $r$, and $\mathcal{R}$ are scaling dimension, $SU(2)_L$ angular momentum, $U(1)_{\mathcal{R}}$ charge, and $SU(2)_{\mathcal{R}}$ charge respectively. The BPS bound is given by $\Delta-2j-2\mathcal{R}-r=0$.

Since we are considering flavor singlet sector of $SU(N)$ QCD with $SU(2N_f)$ flavor symmetry, the theory can be thought of as a quiver gauge theory with $SU(N)\times SU(2N)$ gauge group, where the vector multiplet lives only in $SU(N)$. We consider the index for the trivial vacuum sector such that $SU(N)$ gauge symmetry is unbroken.

It follows that adjoint single letter indices for the quiver matrix model are $f_{11}=t^2v-t^4/v+t^6, f_{22}=0$, and bifundamental letters are $f_{12}=f_{21}=t^2v-t^4/v$.
Here $f_{ii}$ is the single letter index for $i$th gauge node and $f_{ij}$ is the single letter index for bifundamental matters that connects $i$ and $j$th gauge nodes.
One can compute a large $N$ index for quiver-gauge theories if single letter index is given \cite{Gadde:2010en}.
In two gauge-node case, the strict large $N$ index is given by \cite{Gaiotto:2021xce}
\begin{align}\label{index}
    I_{\infty,\infty}=\prod_{n=1}^{\infty}\frac{1}{1-f_{11}-f_{22}+f_{11} f_{22}-f_{12}f_{21}}
\end{align}
where $I_{\infty,\infty}$ denotes the limit in which the ranks of both gauge nodes are taken to infinity while keeping $N_f/N=2$ fixed.
Following this, the index is given by
\begin{align}
    I_{\infty,\infty}=\frac{(1-t^6)(1-t^2v)}{(1-t^4v^{-1})}\prod_{n}\frac{1-t^{6n}}{(1-t^{2n}v^n)(1-t^{12n})}=\text{PE}\left[\frac{t^2v}{1-t^2v}+\frac{t^{12}}{1-t^{12}}-\frac{t^{6}}{1-t^{6}}-f_{11}\right].
\end{align}
As is immediately apparent, the BMN index does not exhibit Hagedorn growth.

This suggests that the matrix quantum mechanics may be viewed as describing a background geometry that is a \textit{generalized} pp-wave, whose isometries include $SU(2)_{\mathcal{R}}\times U(1)_r$, inherited from the $\mathcal{R}$-symmetry. In this interpretation, the index captures the spectrum of supergraviton states propagating in this background. It would be interesting to identify, if any, a corresponding dual geometry for the matrix model; we leave this question for future investigation.

Let us note the universal classical vacuum structure: $X^a\sim J^a$ remains a vacuum in any matrix quantum mechanics obtained from the BMN truncation of four-dimensional supersymmetric gauge theories. This universality follows from the fact that the four-dimensional $\mathcal{N}=1$ vector multiplet is always in the adjoint representation.

\subsubsection{$\mathcal{N}=1$ Seiberg dual theories}

Let us consider a second example, namely $\mathcal{N}=1$ SQCD and its Seiberg-dual description within the conformal window, $\tfrac{3}{2}N \leq N_f \leq 3N$ \cite{Seiberg:1994pq}. The electric and magnetic theories are infrared dual to each other. As a nontrivial check of this duality, one may compute the flavor-singlet superconformal index in the Veneziano limit.
\[
N_c,N_f\to\infty \quad \text{with} \quad r \equiv 1-\frac{N_c}{N_f} \ \text{fixed}.
\]

For a 4d $\mathcal{N}=1$ theory, the superconformal index is defined as
\begin{align}
    I(p,q)=\Tr (-1)^F p^{r+ j_1}q^{r + j_2}
\end{align}
where the fugacities $p$ and $q$ keep track of the Cartan generators of the four-dimensional superconformal algebra. $j_1$ and $j_2$ denote the angular momenta of the $SU(2)_1 \times SU(2)_2$ Lorentz symmetry. 

In the Veneziano limit, the index exponentiates and takes the universal form
\begin{align}
Z_{\infty,\infty}
&=\prod_{n=1}^{\infty}\frac{1}{1-f(p^n,q^n)} ,
\end{align}
where $f(p,q)$ is the single-letter index projected onto the flavor-singlet sector.

For the electric theory, the function $f(p,q)$ can be written in a similar way as in \eqref{index}, where
\begin{align}
1-f(p,q)
&=1-\frac{2 p q-p-q}{(1-p)(1-q)}
-\frac{\left((p q)^{r/2}-(p q)^{1-\frac{r}{2}}\right)^2}{(1-p)^2(1-q)^2},
\end{align}
where the parameter $r$ is the superconformal $R$-charge of the quark superfields,
\[
\frac{1}{3}\leq r = 1-\frac{N_c}{N_f}\leq \frac{2}{3},
\]
as fixed by $a$-maximization in the conformal window.\footnote{Let us note that $r$ is not fixed in the truncated theory because the quantum mechanics is not conformally invariant. However, it is natural to choose $r$, in order to reflect the isometry of the IR geometry.}
More explicitly, 
\[
f_{11}=\frac{2 p q-p-q}{(1-p)(1-q)}, \qquad
f_{22}=0, \qquad
f_{12}=f_{21}=\frac{(p q)^{r/2}-(p q)^{1-\frac{r}{2}}}{(1-p)(1-q)}.
\]
Here $f_{11}$ corresponds to the contribution of the vector multiplet in the adjoint of the gauge group, while $f_{12}$ and $f_{21}$ encode the contributions of the fundamental quarks and antiquarks. The absence of a $f_{22}$ term reflects the fact that there is no adjoint matter multiplet in the electric description.

The expression for $Z_{\infty,\infty}$ of the magnetic theory is the same once the duality map for $r$ is imposed, providing supporting evidence for the IR equivalence of the two theories at large $N_c$ and $N_f$.
As in the previous subsubsection, the superconformal index exhibits Hagedorn growth. For instance, when $p=q$ and $r=\tfrac{1}{2}$, the Hagedorn temperature is reached at $p=\frac{1}{2}(-1+\sqrt{5})$.

On the other hand, the finite-\(N\) BMN truncations obtained from the electric and magnetic descriptions differ from one another.
Since the BMN truncation is not a fully consistent quantum truncation, there is no a priori reason to expect them to coincide.
One indication of this mismatch is that the BMN indices of the electric and magnetic theories should in general differ, except at the special point \(N_f = 2N_c\).\footnote{
The case \(N_f = 2N_c\) is exceptional because the single-letter indices agree in the two descriptions. Even in this case, however, we do not expect the two BMN-like quantum mechanical systems to be dual to each other, although we have not examined this question in detail.}

This suggests that IR-dual four-dimensional theories generally lead to distinct BMN truncations. In other words, if an M-theory dual description of a given BMN truncation exists, it appears to retain sensitivity to UV data of the parent field theory. It would therefore be interesting to understand the origin of this difference and, if so, why the holographic examples with $U(N)$ gauge groups discussed in Subsection \ref{holo} constitute a special class.

\subsubsection{Comment on $SO/Sp$ BMN}\label{sosp}

In this subsubsection we consider matrix models with gauge groups other than \(U(N)\).

One may begin with four-dimensional \(\mathcal{N}=4\) SYM with orthogonal or symplectic gauge groups and perform the usual dimensional reduction to one dimension. This yields BMN-type matrix quantum mechanics with \(SO\) or \(Sp\) gauge groups. At the level of four-dimensional field theory, S-duality relates certain pairs of global forms of the gauge group (for example algebras of type \(B_N\) and \(C_N\) are exchanged under S-duality, with appropriate choices of global form and discrete theta parameter). Thus \(SO(2N+1)\) and \(Sp(N)\) \(\mathcal N=4\) theories are related by S-duality in four dimensions (up to these global data), and their holographic duals involve quotients of the usual \(\mathrm{AdS}_5\times S^5\) background.

However, dimensional reduction, and in particular the BMN truncation, need not commute with duality. Protected quantities computed in the one-dimensional truncations -- for example supersymmetric indices or Witten indices -- differ for the \(SO\) and \(Sp\) matrix models at finite $N$ \cite{Gadde:2025yoa}. This indicates that the BMN truncations of S-dual four-dimensional theories are not equivalent as quantum mechanical models.

The distinction becomes more transparent in the BFSS limit \(\mu\to0\). In that limit the models describe D0-brane quantum mechanics in type IIA string theory on flat space, but now in the presence of an orientifold O0-plane. An O0 orientifold implements the inversion
\[
\vec x \;\mapsto\; -\vec x
\]
on the nine transverse coordinates and admits a discrete choice of RR torsion. Concretely, the orientifold background carries a \(\mathbb Z_2\) torsion class measured schematically by
\[
\int_{\mathbb{RP}^2} F_2 \;=\; 0 \quad \text{or} \quad \tfrac12 \pmod{1},
\]
and the two possibilities lead to different Chan--Paton projections (symmetric versus antisymmetric) and hence to orthogonal or symplectic gauge groups on the D0 worldvolume. Equivalently, the O0-plane comes in two variants, often denoted \(O0^-\) and \(O0^+\), carrying opposite fractional D0-brane charge. This discrete datum makes the corresponding BFSS matrix models physically inequivalent.

Because the BMN plane-wave background is a mass deformation of the BFSS setup, the orientifold choice and its associated discrete RR data survive the deformation: the orientifold action extends to the plane-wave geometry and continues to distinguish the two variants. From the M-theory viewpoint the difference is reflected in the eleven-dimensional lift, where the discrete torsion corresponds to a choice of flux or quotient geometry. Thus the geometric/topological difference between the \(SO\) and \(Sp\) cases explains why their BMN and BFSS truncations are not equivalent, despite the S-duality relation of the parent four-dimensional theories.

In a schematic sense (up to the choice of discrete torsion data), we expect the \(SO/Sp\) matrix models to be dual to a pp-wave background subject to the coordinate identification
\begin{align}
    (x_1,\ldots,x_9)\sim(-x_1,\ldots,-x_9).
\end{align}
A detailed analysis of this correspondence is left for future work.

\section{Discussion}\label{discussion}

In this paper we propose an infinite class of matrix quantum-mechanical models as candidate nonperturbative descriptions of the DLCQ of M-theory on pp-wave–like backgrounds.
Most of the evidence we have presented is symmetry-based. It would therefore be desirable to produce dynamical checks that either support or falsify the proposal and to clarify the mechanism (if any) that makes the reduction work.

Beyond holographic theories, one can apply the same reduction to a wide class of four-dimensional gauge theories. This raises the question of whether some of these models might admit M-theory duals with an effective supergravity description, or at least possess other interesting geometric or physical interpretations.

From the four-dimensional field-theory viewpoint, the angular directions of the six transverse dimensions in the pp-wave background play the role of an “internal’’ manifold together with a radial direction. The isometries acting on the $x_1,x_2,x_3$ coordinates fill out an $SU(2)$ subgroup of $SO(4)$, the latter being the isometry group of the $S^3$ on which the four-dimensional field theory is defined.
The structure associated with the $x_1,x_2,x_3$ sector is universal in a kinematic sense across $SU(N)$ supersymmetric gauge theories: the four-dimensional $\mathcal{N}=1$ gauge multiplet is common to all such theories, and this common sector admits the same classical vacuum structure discussed around \eqref{vacua 2}. For this reason the proposed reduction picks out a universal vacuum structure that is largely determined by symmetry and the gauge-multiplet content, rather than by the detailed matter content of each quiver.

It is also interesting to note that recent work has identified structural links between the BMN matrix model and three-dimensional M2-brane theories. For example, Chang \cite{Chang:2024lkw} observed that the Witten index of the BMN model in the irreducible vacuum sector is related, in an appropriate large-$N$ limit, to the superconformal index of ABJM theory \cite{Aharony:2008ug,Kim:2009wb}. More recently, the authors of \cite{Behan:2025hbx} identified a truncation of weakly coupled ABJM in which the action of the one-loop supercharge matches the BMN subsector of $\mathcal{N}=4$ SYM, allowing an infinite tower of fortuitous representatives to be carried across the two theories. These results point to a degree of structural commonality between the frameworks. In a related but distinct direction, since the Penrose limit of AdS$_4\times S^7$ gives the pp-wave background, we expect that an appropriate limit of the ABJM index should be equivalent to the index of the BMN matrix model; it would be interesting to verify this correspondence.

Finally, extending subsection \ref{susy bh}, we plan to study the indices and cohomologies of matrix models in greater detail, with the aim of better understanding black objects in pp-wave–like backgrounds.
The matrix model can also provide insights into the spectral structure of the four-dimensional parent theories, since the BMN truncation allows one to analyze classical cohomologies within a subsector of the full cohomological structure.
It is possible that the spectrum of classical BPS cohomologies in superconformal field theories remains unchanged even at strong coupling, motivating a program to identify classical BPS cohomologies that correspond to those of BPS black holes.\footnote{Recently, counterexamples to this non-renormalization have been found in $\mathcal{N}=4$ SYM with $SO$ gauge groups \cite{Chang:2025mqp,Choi:2025bhi}; however, the $SU$ and $Sp$ gauge groups have not yet been examined.}
The black hole cohomology problem was initiated in \cite{Chang:2022mjp, Choi:2022caq}, and the BMN truncation has proven particularly useful in this context \cite{Choi:2023vdm, Chang:2024lkw, deMelloKoch:2024pcs, Gadde:2025yoa, Gaikwad:2025ugk}.

\section*{Acknowledgments}
We would like to thank Jaehyeok Choi, Shota Komatsu, Henry Lin, Shiraz Minwalla and especially Seok Kim for valuable discussions. The work of E.L. was supported by the Infosys Endowment for the study of the Quantum Structure of Spacetime.

\bibliography{biblio}

\end{document}